\documentclass[]{aastex631}

\shorttitle{$\copyright$ 2024. All rights reserved.} 
\shortauthors{}
\graphicspath{{./}{figures/}}

\begin{document}

\title{Mapping "Brain Terrain" Regions on Mars using Deep Learning}

\author[0000-0002-5785-9073]{Kyle A. Pearson}
\affiliation{Jet Propulsion Laboratory, California Institute of Technology, Pasadena, CA 91125 USA}

\author{Eldar Noe}
\affiliation{Planetary Science Institute, Tucson, AZ 85719 USA}

\author{Daniel Zhao}
\affiliation{California Institute of Technology, Pasadena, CA 91125 USA}

\author{Alphan Altinok}
\affiliation{Jet Propulsion Laboratory, California Institute of Technology, Pasadena, CA 91125 USA}

\author{Alexander M. Morgan}
\affiliation{Planetary Science Institute, Tucson, AZ 85719 USA}



\begin{abstract}

One of the main objectives of the Mars Exploration Program is to search for evidence of past or current life on the planet. To achieve this, Mars exploration has been focusing on regions that may have liquid or frozen water. A set of critical areas may have seen cycles of ice thawing in the relatively recent past in response to periodic changes in the obliquity of Mars. In this work, we use convolutional neural networks \textcolor{black}{(CNN)} to detect surface regions containing ``Brain terrain'', a landform on Mars whose similarity in morphology and scale to sorted stone circles on Earth suggests that it may have formed as a consequence of freeze/thaw cycles. We use large images ($\sim$100-1000 megapixels) from the Mars Reconnaissance Orbiter to search for these landforms at resolutions close to a few tens of centimeters per pixel ($\sim$25--50 cm). Over 58,000 images ($\sim$28 TB) were searched ($\sim$5\% of the Martian surface) where we found detections in 201 images. To expedite the processing we leverage a classifier network (prior to segmentation) in the Fourier domain that can take advantage of JPEG compression by leveraging blocks of coefficients from a discrete cosine transform in lieu of decoding the entire image at the full spatial resolution. The hybrid pipeline approach maintains $\sim$93$\%$ accuracy while cutting down on $\sim$95$\%$ of the total processing time compared to running the segmentation network at the full resolution on every image.

\end{abstract}




\section{Introduction} \label{sec:intro}

Brain terrain is a geologically young terrain that has the potential to enhance our understanding on the role of water in the recent geological history of Mars. “Brain terrain” is a descriptive term given to a decameter-scale surface texture on Mars that consists of labyrinthine ridges and troughs occurring in flat terrains, typically in topographic lows. The areas are named for their resemblance to the human brain or aquatic brain coral species.  It is found primarily at mid-latitudes and is often associated with lineated valley fill (LVF), concentric crater fill (CCF) and lobate debris aprons (LDA) (e.g., \citealt{Squyres1978}; \citealt{Malin2001}; \citealt{Carr2001}; \citealt{Mangold2003}). This terrain type shares morphological similarities to sorted stone circles on Earth, which are thought to form as the result of numerous freeze/thaw cycles of rock-bearing soil \citep{Dobrea2007}. In Earth’s arctic environments, sorted stone circles and labyrinths are the result of a rock-bearing and water-rich soil layer that undergoes heave/contraction cycles as the result of freezing-thaw processes (\citealt{Taber1929}; \citealt{Taber1930}; \citealt{Williams1989}) whereby rock-soil segregation can systematically occur (\citealt{Konrad1980}) and lead to pattern formation (e.g. \citealt{Werner1999}; \citealt{Kessler2003}). The convective kinematics of this process are reasonably well explored (\citealt{Goldthwait1976}; \citealt{Williams1989}; \citealt{Hallet2013}) through the analysis of trenches and seasonal data from tilt meters and other field studies (e.g. \citealt{Hallet2020}). However, while the morphologies of brain terrain and sorted stone circles are similar, similarity in form does not necessarily imply the same underlying process. Competing hypotheses argue brain terrain formed by sublimation lag, polygon inversion, or stone sorting by freeze-thaw (\citealt{Mangold2003}; \citealt{Dobrea2007}; \citealt{Levy2009}), but its origin remains unresolved. Multiple classes of young geomorphic features on Mars suggest that thawing may have occurred in the recent geologic past and may still be ongoing, although significant controversy remains \cite{McEwen2011}. It is therefore important to compare available hypotheses (\citealt{Mangold2003}, \citealt{Milliken2003}, \citealt{Costard2002}, \citealt{Kreslavsky2008}, \citealt{Cheng2021}, \citealt{Hibbard2022}) by performing a careful and detailed study of this terrain. 

Sorted stone circles on Earth are thought to form from cyclic freezing and thawing in permafrost regions, which drives the convection of stones, soil, and water in the active layer \citep{Mangold2005}. In western Spitsbergen, the stone circles consist of a central 2-3 m wide plug of soil surrounded by a 0.5-1 m wide ring of stones, with the plug domed up to 0.1 m and the stones rising up to 0.5 m \citep{Kessler2001}. Though stones are often cm-scale, meter-sized clasts in circles up to 20 m across occur as well (\citealt{Trombotto2000}; \citealt{Balme2009}). Comparatively, we measure brain terrain cells on Mars to be $\sim$5-15 m with ridges 0.2-2 m tall, resembling terrestrial circles in scale (\citealt{Kessler2001}; \citealt{Kaab2014}). On Mars, brain terrain is relatively young based on crater counts (Morgan et al. \textit{in prep}), though rigorous analysis is still needed. 

The Mars Reconnaissance Orbiter (MRO) has collected images of the Martian surface for over 15 years and amassed over 50 terabytes of data. The High Resolution Imaging Science Experiment (HiRISE; 0.3 m/pixel resolution; \citealt{McEwen2007}) and the Context Camera (CTX; 6 m/pixel resolution; \citealt{Malin2001}) are two instruments onboard MRO that are routinely used to study geological landforms. The total volume of data from this mission poses new challenges for the planetary remote-sensing community. For example, each image includes limited metadata about its content, and it is time-consuming to manually analyze each image by eye. Therefore, there is a need for computational techniques to search the HiRISE and CTX image databases and discover new content.

Many algorithms can classify image content, often requiring preprocessing like edge detection or histogram of oriented gradients \citep{Dalal2005}. \textcolor{black}{For pattern recognition tasks like object classification, convolutional neural networks (CNNs) have emerged as a powerful approach and} can now match human performance in object recognition tasks making them an excellent choice for an algorithm capable of learning optimal features from training data (\citealt{He2015}; \citealt{Ioffe2015}). \textcolor{black}{CNNs are particularly well-suited for image classification tasks because of their ability to learn optimal filters or feature detectors from the training data itself, eliminating the need for hand-crafted features or preprocessing steps like edge detection or histogram of oriented gradients.} For planetary mapping, CNNs outperform other classifiers like support vector machines \citep{Palafox2017}, and can be fine-tuned to identify landforms in Martian images (\citealt{Wagstaff2018}; \citealt{Nagle2020}). Other relevant approaches for planetary exploration include terrain segmentation to inform navigation \citep{Dai2022} and caption generation for planetary images \citep{QIU2020}. \textcolor{black}{To further improve computational efficiency, especially when dealing with large image archives, recent work has explored leveraging the frequency domain representations of images. Leveraging frequency domain data saves significant time when processing large archives by avoiding full decoding. This enables timely surveys to inform landing site selection and identify rare features like potential astrobiology targets. Image compression relies on frequency domain transforms like discrete cosine transforms in JPEG to remove redundant data. Recently, deep learning approaches have utilized these compressed representations for efficient near state-of-the-art object classification (\citealt{Russakovsky2015}; \citealt{Gueguen2018}; \citealt{Chamain2019}). }

We are interested in automating the detection of brain terrain with images from MRO/HiRISE \textcolor{black}{because their resemblance to terrestrial periglacial features makes them promising regions to further investigate in the search for potential life on Mars, both past and present.} To expedite the processing we leverage a classifier network (prior to segmentation) in the Fourier domain that can take advantage of JPEG compression by leveraging blocks of coefficients from a discrete cosine transform in lieu of decoding the entire image at the full spatial resolution. The hybrid pipeline approach cuts down on the total processing time compared to running a segmentation network at the full resolution on every image. In the sections below we will discuss how the training data is made, what goes into each of the networks we tested, and then quantify the performance and accuracy of predicting brain terrain.

\section{Observations and Training Data}

\begin{figure}[h]
\hspace{-0.25in} \\
\includegraphics[scale=0.35]{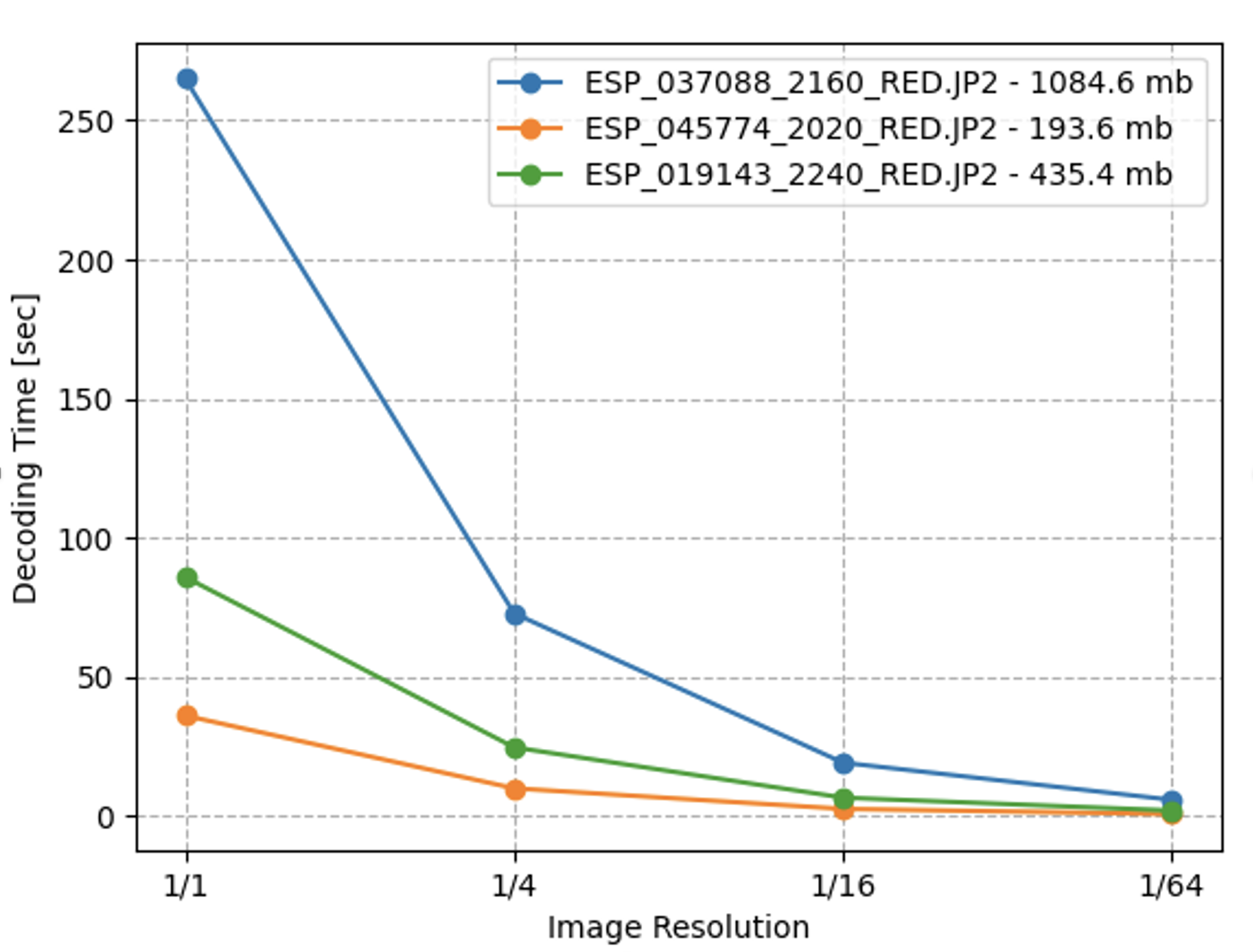}
\caption{ Decoding times of various JPEG2000 images in the HiRISE archive using a single thread on a 2.4 GHz 8-Core Intel Core i9 CPU. The decoding time improves if the image is opened at a lower resolution. However, a lower resolution may not always retain enough signal to noise for a computer vision algorithm to act reliably. The image size on disk is listed in the plot legend. }
\label{fig:decoding_time}
\end{figure}

We analyze observations of the Martian surface using the High Resolution Imaging Science Experiment onboard the Mars Reconnaissance Orbiter (MRO) \citep{McEwen2007}. Images from the primary and extended mission were used and are publicly available online\footnote{https://hirise.lpl.arizona.edu/}. HiRISE has a 0.5 m telescope along with an optical camera however for the purposes of our study we use images from the red channel which has a photometric range between $\sim$550--800 nm. The MRO spacecraft orbits Mars at an altitude of around 300 km and depending on the pointing, the HiRISE camera is capable of imaging the surface at resolutions between $\sim$25-50 cm/px. The camera reads swaths of data at a time tracking the ground in spans of 20,000 pixels and combined images are usually 100,000's pixels  long. The vast amount of image data is stored with 10-bit precision as a JPEG2000 image. The image encoding algorithm is capable of compressing images to $\sim$5-10$\%$ their original size, e.g., 10.8 GB of data at the full spatial resolution ($ESP\_036917\_2210$) compresses into as little as $\sim$675 MB of space on disk. Decoding images at the full-resolution can often take up to a few minutes on a CPU depending on the image size (see Figure \ref{fig:decoding_time}). \textcolor{black}{The HiRISE archive contains over 50 TB of compressed data spanning $\sim$85,000 images (as of writing this paper). However, for our study we used a local snap shot acquired in 2020 with 58,209 images.} Processing terabytes worth of images requires spending a non-negligible time decoding the data alongside applying any computer vision algorithm afterward.

In order to automate the detection of landforms on the Martian surface we need a computer vision algorithm capable of learning the ideal features for detection. The advantage of a neural network is that it can be trained to identify subtle features \textcolor{black}{(e.g. in pixel or elevation data)} inherent in a large data set. This learning capability is accomplished by allowing the weights and biases to vary in such a way as to minimize the difference (i.e. the cross entropy) between the output of the neural network and the expected or desired value from the training data. Our training data is separated into two classes; a generic `background` class that represents most types of land and a `brain terrain` class for our terrain of interest. The training data is split in a 3:1 ratio in order to provide a diverse range of background samples. \textcolor{black}{This skewed distribution prevents bias towards the majority class and provides diverse negative samples, enabling effective discrimination of brain terrain from visually similar terrains (e.g. aeolian ridges or deflationary terrain) while minimizing false positives. We tested different ratios and found a bigger ratio (more background) created a higher rate of false negatives while smaller ratios yielded more false positives, including a 1:1 split. Crucially, this imbalanced scenario simulates learning from limited labeled examples, a common challenge when studying rare phenomena, thus building robustness for deployment on new data where brain terrain occurrences are sparse yet scientifically compelling across the vast Martian surface. It prevents the model from becoming biased towards the overrepresented majority class and instead forces it to learn highly discriminative features for precise identification. }

When we started this study, we only had two hand-labeled examples of brain terrain and used active learning to build a more robust set of training data.  \textcolor{black}{The initial two hand-labeled examples of brain terrain came from the images published in \citealt{Dobrea2007}. We used those labeled regions to train an initial classifier to identify similar structures in other HiRISE images. We then manually vetted the classifier's detections, adding any false positives to the background training set and any missed detections of true brain terrain to the positive training set. This active learning process of training, manually vetting, and adding to the training sets was iterated on to steadily improve the classifier's performance. Each new image added to our brain terrain training data was masked manually to avoid biasing based on outputs from a preliminary algorithm.} The full list of images used to generate training samples is listed in Table \ref{tab:training}. 

\begin{figure}[h]
\includegraphics[scale=0.6]{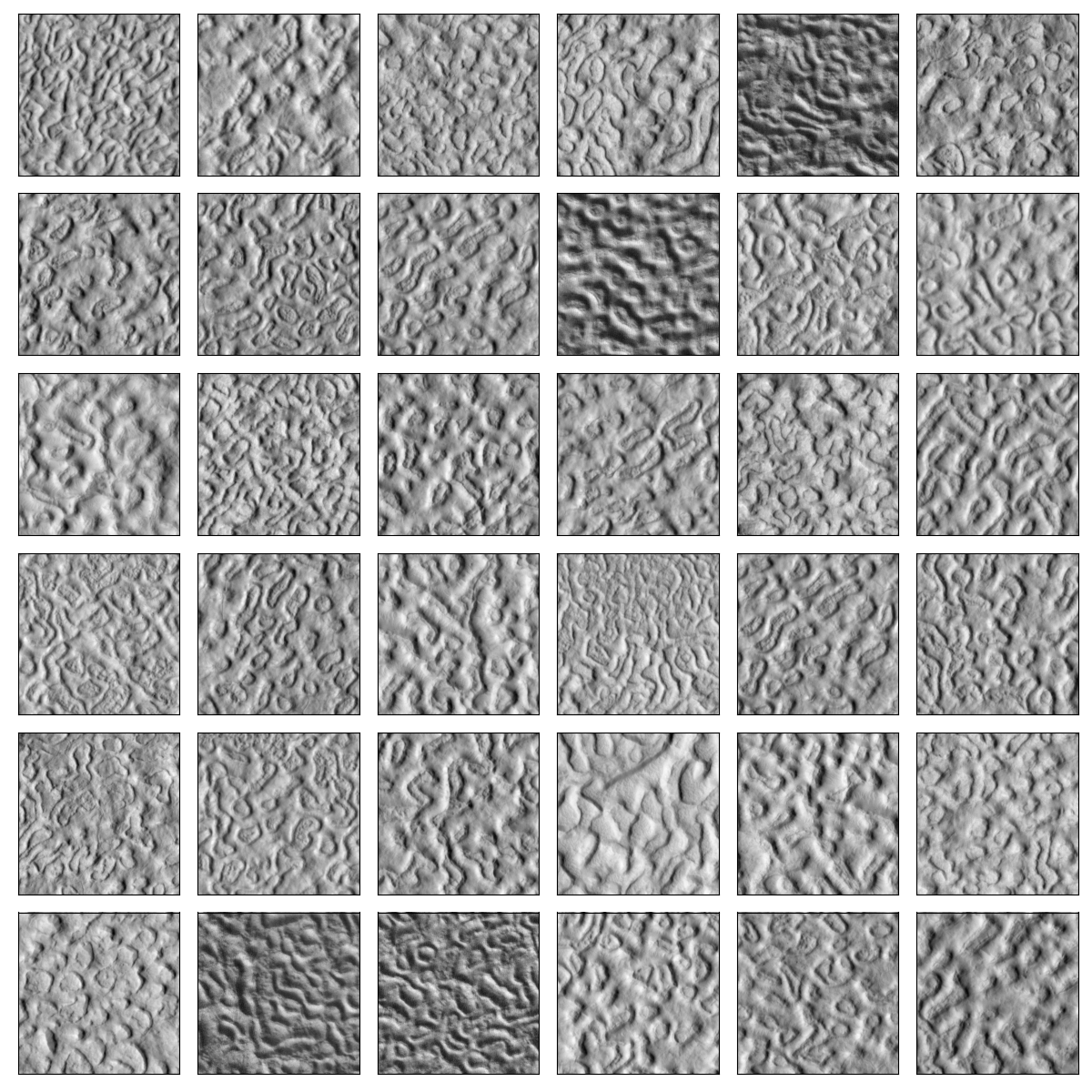}
\caption{ A mosaic of training samples for the brain terrain class. Each tile shown here has a size of 512 $\times$ 512 pixels which corresponds roughly to $\sim$128$\times$128$m^{2}$ area at the native resolution. The tiles shown here are reminiscent of the data used to train the segmentation algorithm. Although, images pertaining to the background class are not shown here. The brain terrain exhibits pitted mounds next to small ridges which range in size from 1-10 meters. Most of the examples shown here are open-cell brain terrain (e.g. ESP$\_$057317$\_$2210 and ESP$\_$061075$\_$2195). Closed cell terrain can be found in PSP$\_$001410$\_$2210. 
}
\label{fig:training_mosaic}
\end{figure}

It is important to train on a diverse population of background images since it needs to generalize to the \textcolor{black}{$\sim$60,000 images in the HiRISE archive snapshot that we looked at for this study}. We trained multiple networks for our study and used the one with the best results. We found shrinking the input window size usually degrades performance below 128 pixels, therefore all of the classifier networks we tested have bigger window sizes (either 128, 256, or 512 pixels; see Table \ref{tab:comparison}). The networks are trained on data between $\sim$25-50 cm/px and supports the conclusion of other papers stating due to the convolutions and pooling operations the network is partially scale invariant (\citealt{Xu2014}; \citealt{Xu2019}; \citealt{Wimmer2023}). Obviously, if the brain terrain feature is larger than the window size this conclusion breaks down and the network needs a larger input window. We generate training tiles using the map-projected satellite images and ignore portions of the image containing the black border. A subset of our brain terrain training samples can be seen in Figure \ref{fig:training_mosaic}.

\begin{deluxetable*}{llllll}
\centering
\tablecaption{Training Data\label{tab:training}}
\tablehead{
\colhead{Name} & Class & Description}
\startdata
ESP$\_$016215$\_$2190  & brain terrain & Area West of Erebus Montes \\
ESP$\_$016287$\_$2205  & brain terrain & Lineated Valley Floor Material in Terrain North of Arabia Region \\
ESP$\_$018707$\_$2205  & brain terrain & Candidate New Impact Site Formed between January 2010 and June 2010 \\
ESP$\_$022629$\_$2170  & brain terrain & Lobate Deposits in Crater in Arabia Terra \\
ESP$\_$042725$\_$2210  & brain terrain & Lineated Valley Floor Material in Terrain North of Arabia Region \\
ESP$\_$052675$\_$2215  & brain terrain & Region Near Erebus Montes \\
ESP$\_$057317$\_$2210  & brain terrain & Crater with Preferential Ejecta Distribution on Possible Glacial Unit \\
ESP$\_$061141$\_$2195  & brain terrain & Candidate Landing Site for SpaceX Starship in Arcadia Region\\
PSP$\_$001426$\_$2200  & brain terrain & Lobate Apron Feature in Deuteronilus Mensae Region \\
PSP$\_$007531$\_$2195  & brain terrain & Striated Flows in Canyon\\
ESP$\_$019385$\_$2210  & brain terrain & Lineated Valley Floor Material in Terrain North of Arabia Region \\
ESP$\_$036917$\_$2210  & brain terrain & Doubly-Terraced Elongated Crater in Arcadia Planitia \\
ESP$\_$052385$\_$2205  & brain terrain & Sample Region Near Erebus Montes\\
ESP$\_$055146$\_$2220  & brain terrain & Candidate Recent Impact Site \\
ESP$\_$060698$\_$2220  & brain terrain & Subliming Ice \\
ESP$\_$061075$\_$2195  & brain terrain & Candidate Landing Site for SpaceX Starship in Arcadia Region \\
ESP$\_$077488$\_$2205  & brain terrain & Lineated Valley Floor Material in Terrain North of Arabia Region \\
PSP$\_$001410$\_$2210  & brain terrain & Impact Crater Filled with Layered Deposits\\
PSP$\_$001466$\_$2215  & brain terrain & Fretted Terrain Valleys and Apron Materials\\
PSP$\_$009740$\_$2200  & brain terrain & Debris Aprons in Eastern Erebus Montes\\
\hline
ESP$\_$011428$\_$2200  & Background & Candidate New Impact Site Formed between January and October 2008 \\
ESP$\_$011571$\_$2270 & Background & Very Fresh Small Impact Crater\\ 
... & ... & ... \\
PSP$\_$008158$\_$1825 & Background & Region North of Nicholson Crater\\
PSP$\_$010346$\_$1570 & Background & Graben in Memnonia Fossae \\ 
\enddata
\tablecomments{The full table consists of 149 images and is hosted online. }
\end{deluxetable*}

\begin{figure}
\hspace{-0.25in} \\
\includegraphics[scale=0.42]{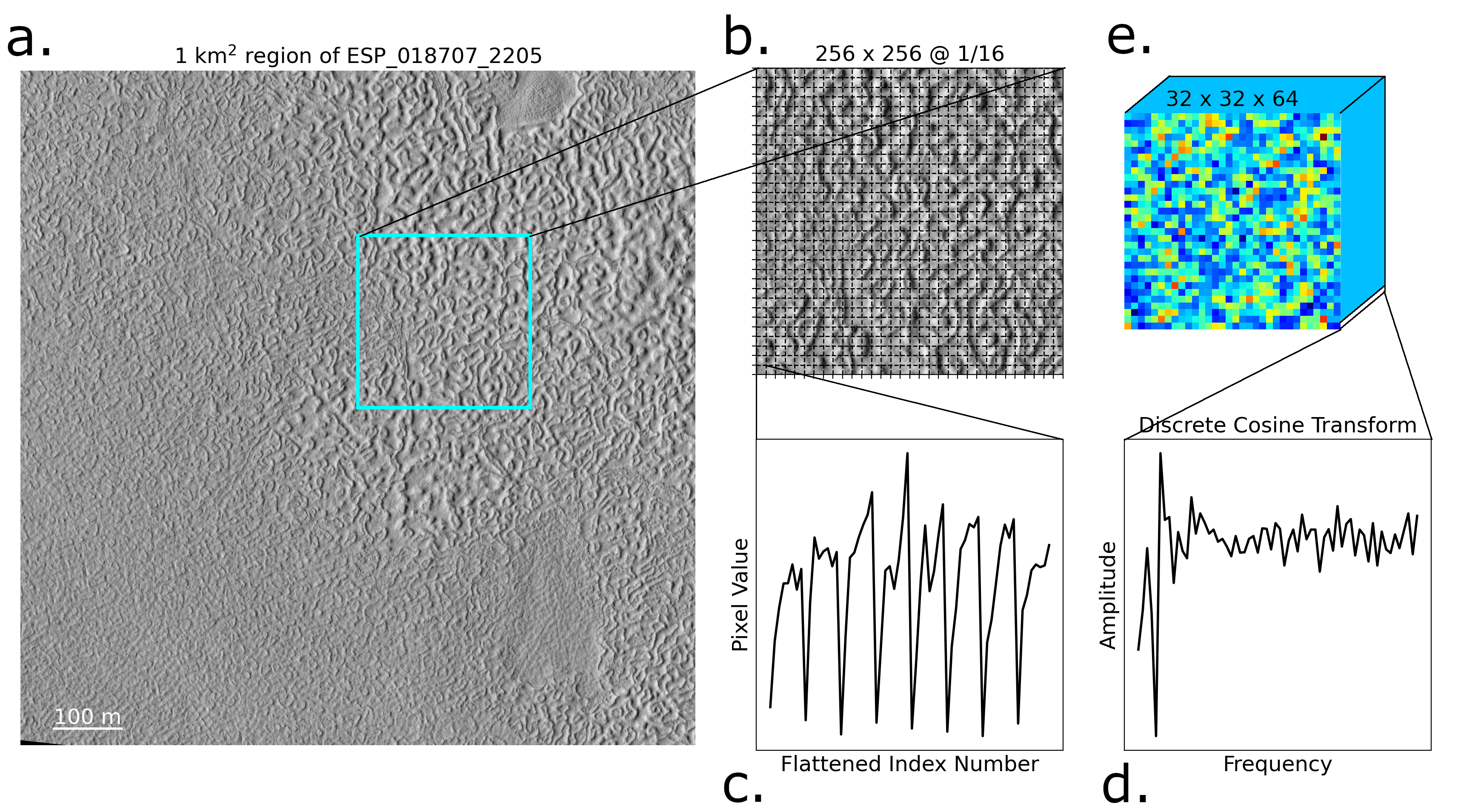}
\caption{ An overview of the inputs for each step in our processing pipeline. \textbf{a)} A region in the HiRISE image of ESP\_018707\_2205 is shown at the native resolution (0.3m/px) with a blue box highlighting the window size for our classifier network. \textbf{b)} A 256 x 256 pixel window is used as input for our spatial classifier algorithm however the window is at 1/16 the original resolution (1.2m/px). A grid of 8x8 squares is displayed showing how that image gets tiled and preprocessed using parts of a JPEG encoder which involves the discrete cosine transform (d). \textbf{c)} A single 8x8 tile flattened into a 1d array which is used for the DCT transform (d). \textbf{e)} A block of Fourier coefficients is rearranged into a data cube and used as input for our Fourier classifier. Reducing the image size and ultimately the channel size after the first conv. layer significantly shortens the network's processing time compared to the spatial image input.}
\label{fig:fourier_domain}
\end{figure}

\section{Cross-Validation and Algorithm Comparison}

A hybrid pipeline approach allows us to save processing time by evaluating data at a lower resolution while maintaining most of the accuracy compared to a full-resolution network. The hybrid approach starts with a classifier for quickly evaluating images followed by a segmentation network for full-resolution pixel-scale inference. Multiple classifiers were tested using different architectures and input sizes which leads to differences in accuracy and evaluation speed (see Table \ref{tab:comparison}). We tested two types of classifiers, one in the Fourier domain in order to leverage blocks of coefficients from an encoding process similar to JPEG and a spatial classifier for comparison. Both classifiers work with data at 1/16 resolution. The spatial classifier uses a normalization layer in the network to transform the input to have a mean of 0 and a standard deviation of 1. The Fourier domain classifier uses the JPEG encoding processing to tile an input into 8x8 tiles and then applies a discrete cosine transform to each tile before being reshaped into a smaller but deeper block for the network (see Figure \ref{fig:fourier_domain}). The tiling and encoding process can transform an input image of 256x256 into a data cube of size 32x32x64. The smaller window size is advantageous in convolutional neural networks since the first layer of the network usually reduces the dimensionality in the number of channels from 64 to 32. Additionally, the smaller image size requires fewer operations per layer and speeds up inference whereas, with a similar architecture for the spatial data, the first layer increases the dimensionality of the input since there is only one channel pertaining to color instead of frequency. Early on in our study we started with random forests and multi-layer perceptrons but quickly found more benefit in using bigger input sizes which are better suited for convolutional networks. We found a negative correlation between the input size and the number of false positives with every algorithm we tested and ultimately settled on window sizes of 128x128 or 256x256, depending on the classifier. The advantage of using a smaller window size like 128x128 is that we can make more training data for a given set of HiRISE images (see Table \ref{tab:comparison}).

\textcolor{black}{
While the classifier network demonstrated high accuracy and computational efficiency, it lacked the precision required for detailed surface measurements. Detecting brain terrain at sub-meter scales necessitated a network capable of operating at the native HiRISE image resolution ($\sim$0.25-0.5 m/pixel). To this end, we designed a custom U-net architecture based on MobileNetV3 to build a segmentation network for pixel-level predictions \citep{Howard2019}. The MobileNet architecture was chosen due to its state-of-the-art accuracy on the CiFAR1000 test while requiring less training time and computational resources compared to larger CNN architectures like ResNet and VGG \citep{He2015}. Despite its name implying mobile applications, MobileNet was designed for ARM-based processors, which are growing in popularity for space-based operations due to their energy efficiency \citep{Dunkel2022}.}

\textcolor{black}{
The U-net architecture is widely adopted for segmentation tasks due to its ability to learn both image context and object boundaries (\citealt{Ronneberger2015}; \citealt{Garcia2017};). It consists of a contracting encoder path that applies convolutions to downsample the input image and extract relevant features, and an expanding decoder path that upsamples these features to produce a pixel-wise segmentation mask at the same resolution as the input. The encoder uses a series of convolutional and pooling layers to downsample the input image, extracting higher-level semantic features at deeper layers. The decoder then upsamples these feature maps to produce full-resolution segmentation masks. Crucially, the decoder combines high-level semantic features from the encoder's deepest layers with higher-resolution features from earlier layers via skip connections. This allows the U-net to integrate local information (from early layers) with global context (from deep layers) to make accurate pixel-wise predictions. Compared to standard convolutional architectures like VGG (a very deep stack of 3x3 convolutional layers), the U-net's decoder path and skip connections enable much more precise segmentation around the boundaries of objects/regions. This precise segmentation is highly beneficial for delineating the intricate patterns of brain terrain at a high spatial resolution. We chose MobileNetV3 as the core encoder architecture due to its efficient design achieving high accuracy at low computational cost, as highlighted in the \citealt{Howard2019}. This allowed fitting a high-resolution U-net within memory constraints.}

\begin{figure}[h]
\hspace{-0.25in} \\
\includegraphics[scale=0.65]{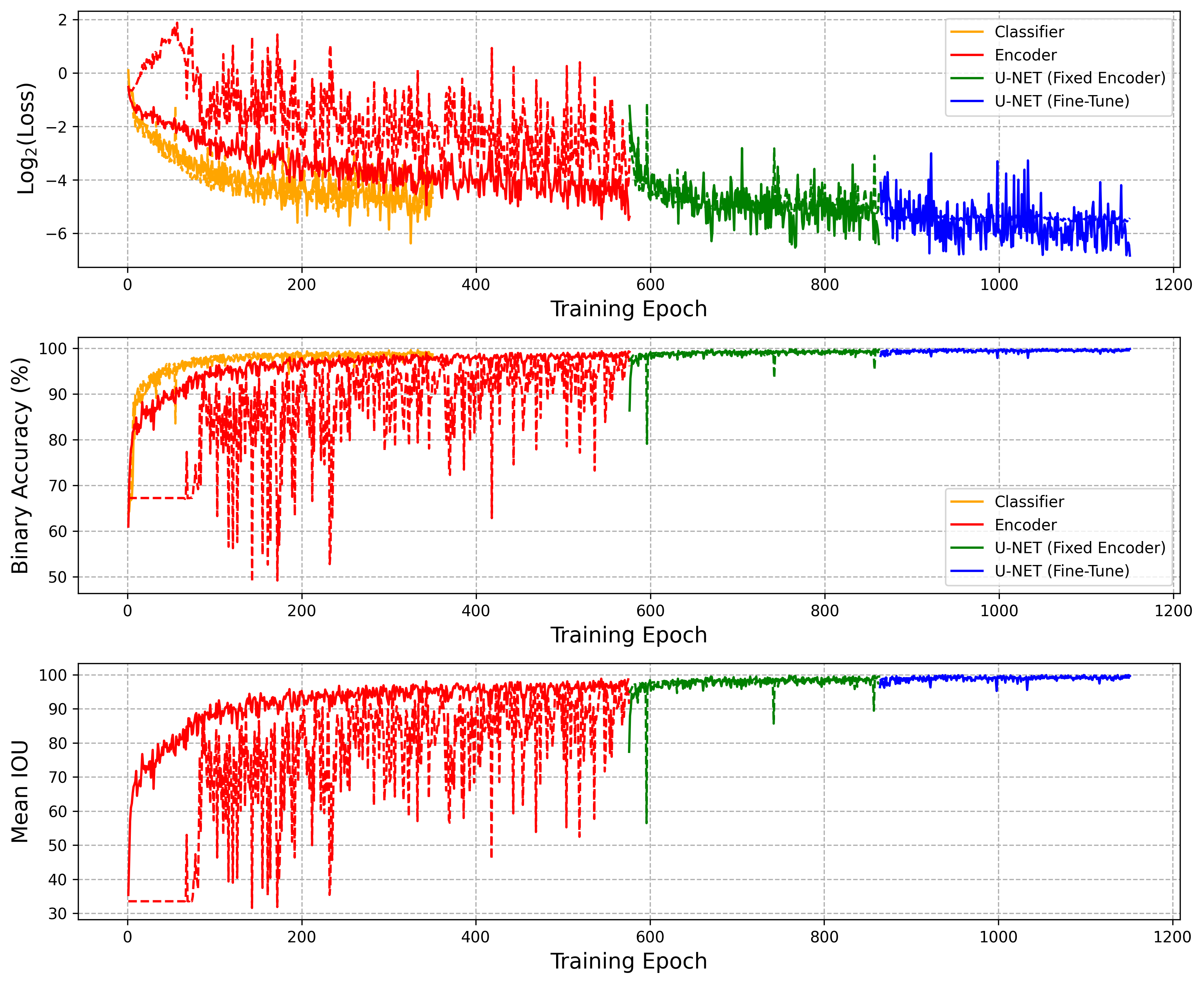}
\caption{ A plot comparing the training behavior for the networks used in our survey. The classifier leverages data in the Fourier domain from a spatial image at 1/16 the original resolution. Where as the U-net algorithm uses spatial data at the native resolution and a MobileNetV3 architecture. We found training the U-Net in pieces yielded better testing performance. }
\label{fig:training}
\end{figure}

\begin{figure}
\hspace{-0.25in} \\
\includegraphics[scale=0.35]{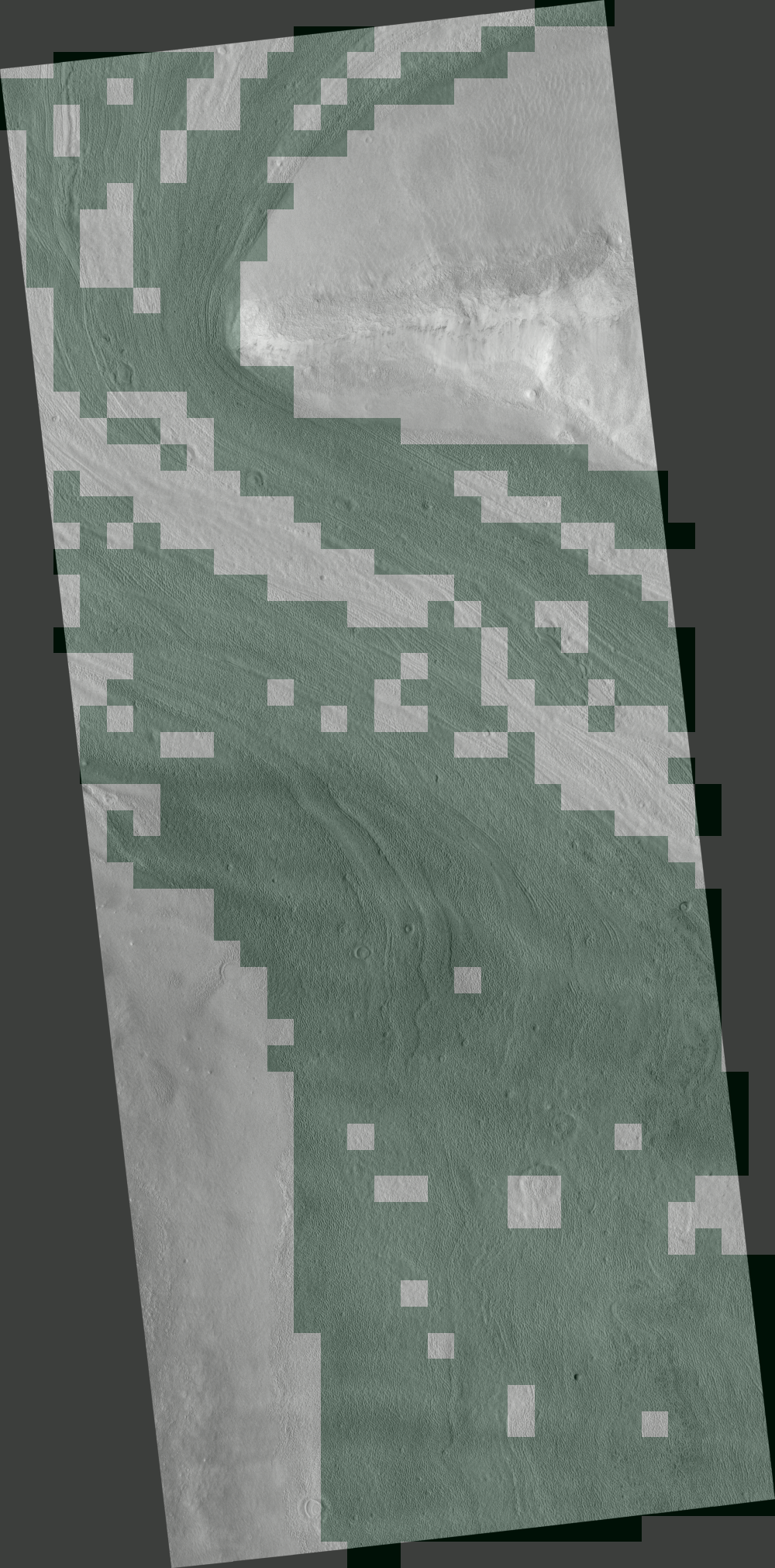}
\includegraphics[scale=0.35]{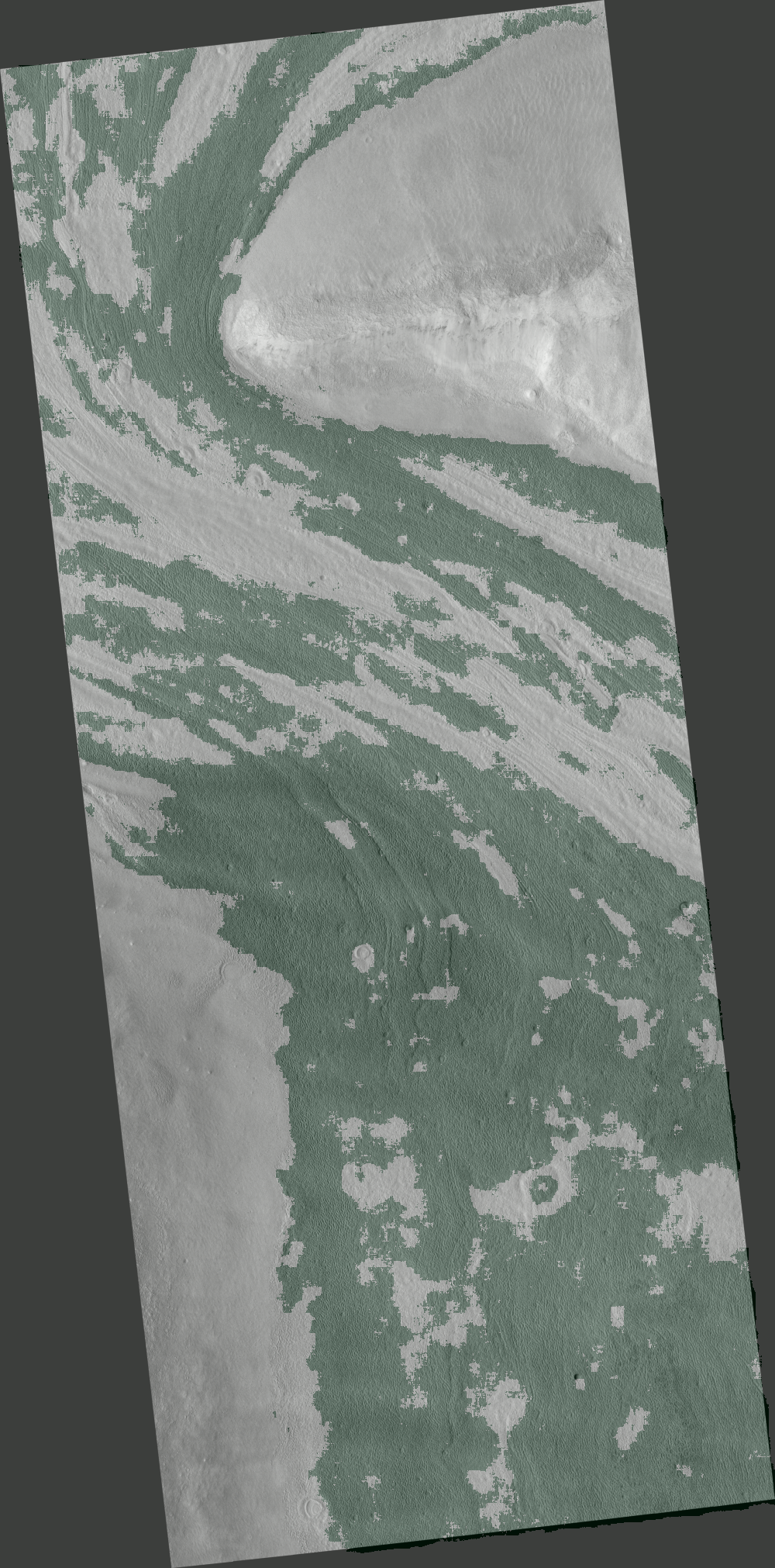}
\caption{ A comparison of outputs between our final classifier and segmentation network overlaid on the image $ESP\_016287\_2205$. \textbf{Left)} An output from our classifier corresponding to detections from the brain terrain class. The pixelated output corresponds to the field of view for the classifier ($\sim$64x64 meter region at 1/16 resolution). \textbf{Right)} A segmentation mask at the native image resolution from the U-net algorithm for the brain terrain class. The brain terrain is highlighted in \textcolor{black}{green} in both images. }
\label{fig:mask_comparison}
\end{figure}

Due to the size and complexity of the U-Net architecture, we found improved performance by training the network in multiple stages. First, the encoder portion of the U-Net was trained to predict low-resolution (32x32) segmentation masks from the input images (512x512). As shown by the jitter in testing accuracy (Figure \ref{fig:training}), the encoder alone did not generalize adequately to the testing data. Next, the full U-Net was assembled by appending the decoder portion and fixing the encoder weights. Only the decoder was trained, and conditioned on upscaling the low-resolution encoder outputs to the native image resolution. Skip connections from select encoder layers to decoder layers of equal depth were utilized, as per the U-Net design. Since the base network architecture (MobileNet) is not a U-Net, determining which layers to use for skip connections required empirical tuning. The number of upsampling layers in the decoder constrained possible skip connection sources in the encoder. Finally, end-to-end fine-tuning of the full U-Net was performed using a learning rate an order of magnitude lower than the previous training stages. The Adam optimizer was used for training, with binary cross-entropy loss, as there were only two classes of interest. A comparison of the classifier and segmentation network is in Figure \ref{fig:mask_comparison}.

\textcolor{black}{
To evaluate the performance of the different machine learning architectures for automating the detection of brain terrain, we used the F1 score as the primary metric. The F1 score provides a holistic measure by combining precision (the fraction of detected regions that are true positives) and recall (the fraction of actual brain terrain regions successfully detected). It ranges from 0 to 1, with a value of 1 representing perfect precision and recall. As shown in Table \ref{tab:comparison}, the unet-512-spatial model achieved the highest F1 score of 0.998 on the test dataset, indicating excellent overall accuracy in correctly identifying brain terrain while minimizing both false positive and false negative detections. The F1 metric enables evaluating the critical trade-off between reducing false alarms and avoiding missed detections for this classification task. Maximizing the F1 score ensures the automated mapping approach has high fidelity for subsequent scientific analysis of the spatial distribution and morphology of brain terrains.}

\begin{deluxetable}{lcclllllc}
\centering
\tablecaption{Network Performance\label{tab:comparison}}
\tablehead{
model name& Training Size& Testing Size& F1 Score& TP& TN& FP& FN& 1Kx1K / sec }
\startdata
unet-512-spatial &  58641& 6516& 0.998 & 99.8 & 99.7 & 0.2 & 0.3 & 6.5 \\
cnn-128-spatial & 58641& 6516& 0.996& 99.6& 99.6& 0.4& 0.4 & 33.9 \\
resnet-128-spatial & 58641& 6516& 0.995& 99.7& 99.4& 0.6& 0.3 & 21.7 \\
cnn-256-spatial & 11493& 1277& 0.991& 99.2& 99.0& 1.0& 0.8 & 46.4 \\
resnet-256-spatial & 13736& 1527& 0.990& 99.6& 98.5& 1.5& 0.4 & 35.1 \\
MobileNet-128-spatial & 58641& 6516& 0.988& 98.6& 99.0& 1.0& 1.4 & 97.7 \\
MobileNet-256-spatial & 13736& 1527& 0.977& 96.0& 99.5& 0.5& 4.0 & 169.0 \\
cnn-256-dct & 13736& 1527& 0.930& 90.6& 95.7& 4.3& 9.4 & 228.4 \\
cnn-128-dct &  58641& 6516& 0.912& 90.4& 92.1& 7.9& 9.6 & 134.2 \\
resnet-256-dct &13736& 1527& 0.844& 87.3& 80.5& 19.5& 12.7 & 130.0 \\
MobileNet-256-dct & 13736& 1527& 0.836& 93.6& 69.6& 30.4& 6.4 & 159.8 \\
\enddata
\tablecomments{All metrics are computed using the test data. The last column represents how many 1K x 1K windows each model can process per second on a RTX 3090 GPU. The numbers in the model name represent the size of the input window in pixels. TP = True Positive, TN = True Negative, FP = False Positive, FN = False Negative. }
\end{deluxetable}

\begin{figure}[h]
\includegraphics[scale=0.55]{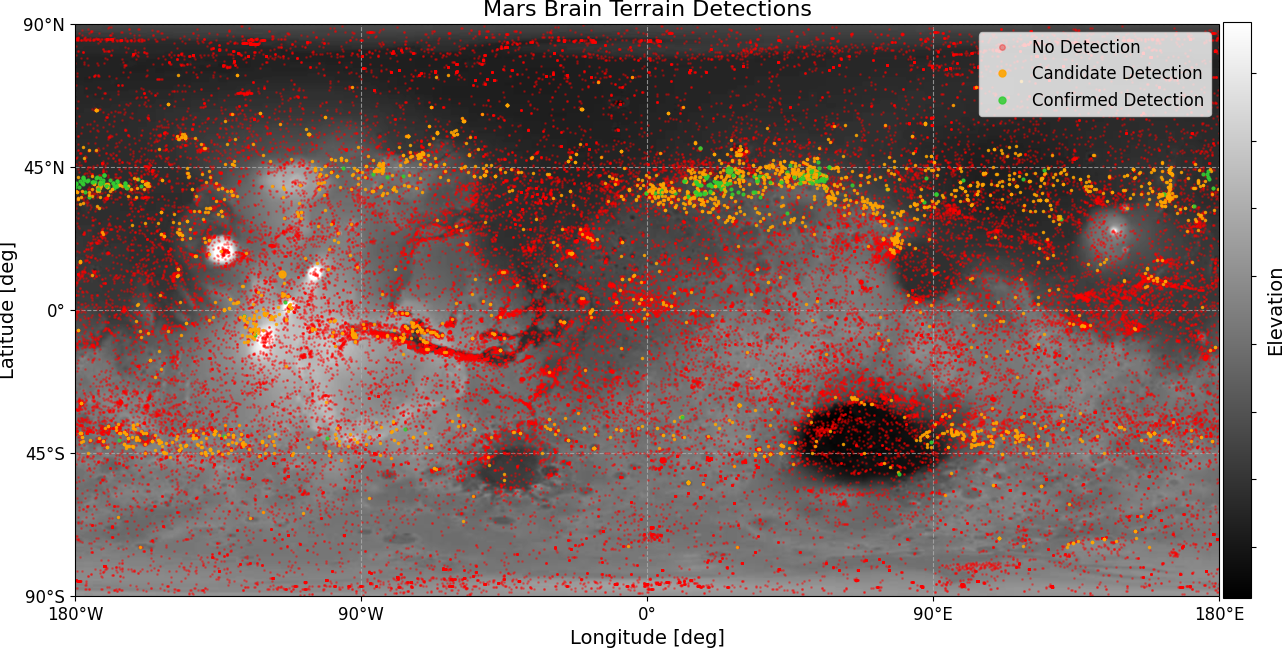}
\includegraphics[scale=0.55]{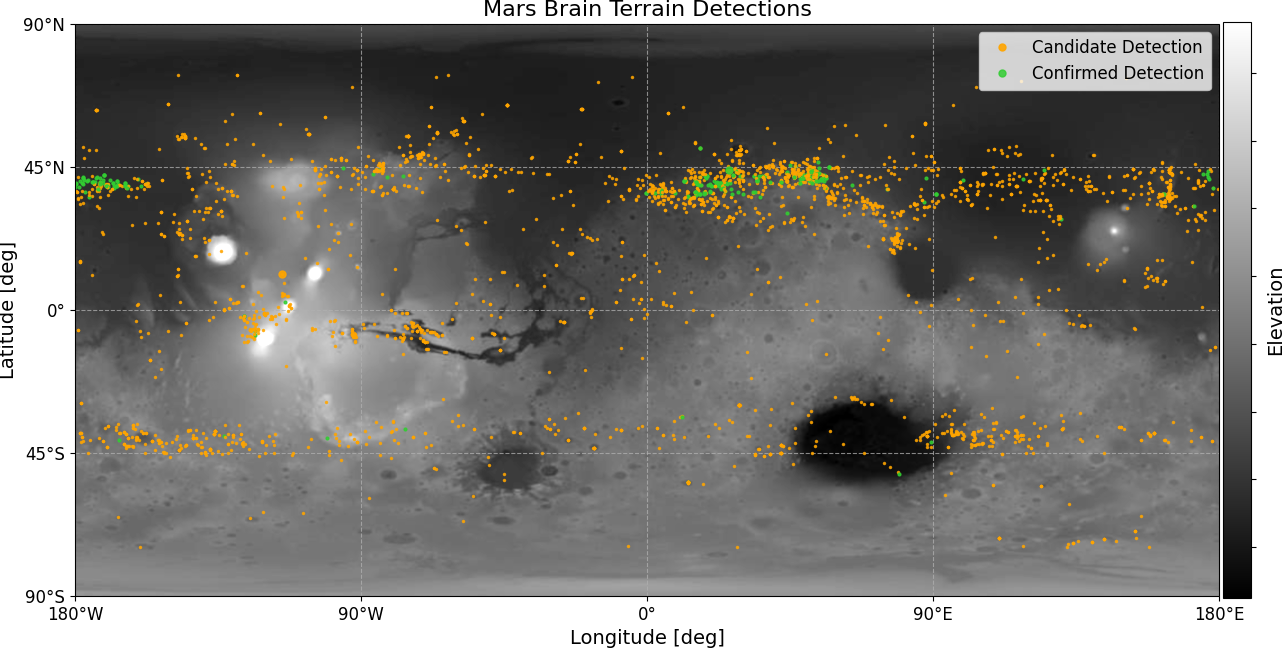}
\caption{ A map of candidate regions that contain brain terrain plotted over elevation data. Our survey vetted 58,209 HiRISE images down to 201 confirmations and 1141 candidates. The orange points are images flagged by the classifier and segmentation algorithm while the green points were once orange but have been manually vetted by our team afterwards. The elevation data is blended from the Mars Orbiter Laser Altimeter (MOLA), an instrument aboard NASA’s Mars Global Surveyor spacecraft (MGS), and the High-Resolution Stereo Camera (HRSC), an instrument aboard the European Space Agency’s Mars Express (MEX) spacecraft. }
\label{fig:mars_map}
\end{figure}

\begin{figure}[h]
\includegraphics[scale=0.6]{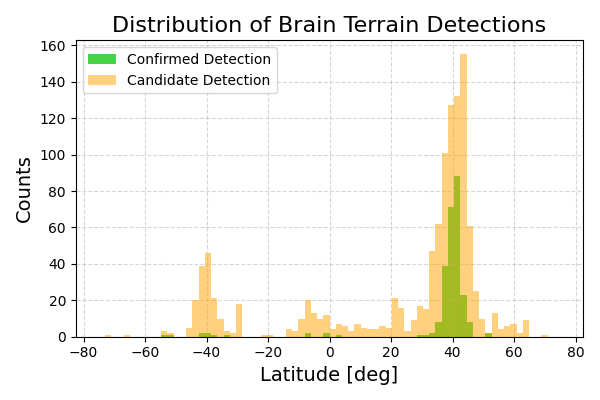}
\includegraphics[scale=0.6]{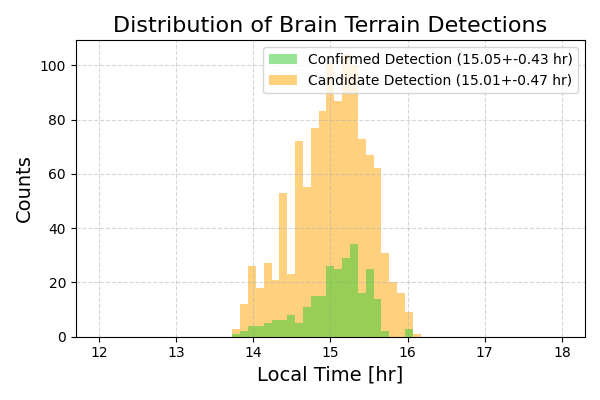}
\caption{ A distribution of regions that contain brain terrain displaying latitude on the left and local time on the right. The confirmed detections were manually checked after our algorithms flagged an image where as the candidate detections have not been manually vetted. The latitude distribution has peak around 40N and another peak around 40S. The preferential distribution in latitude suggests whatever process formed brain terrain also ties to certain climate conditions. Our brain terrain detections and training data span a range of local times suggesting the algorithm is robust against a range of illumination conditions. }
\label{fig:mars_latitude}
\end{figure}


\section{Results and Discussion}

The HiRISE camera on MRO has been acquiring data of the Martian surface for over a decade. \textcolor{black}{We acquired a snapshot of the archive in 2021 which contains 58,209 images subtending over 28.1 TB of space on disk.} We have a problem of needing to find a rather rare terrain over many images and \textcolor{black}{we need pixel-level boundaries in order to accurate measure the spatial extent} (see Figure \ref{fig:mars_map}). Figure \ref{fig:decoding_time} shows the amount of time it takes to decode a single HiRISE image based on the decoding resolution and image size. The estimated time to decode the entire archive at the native resolution is about 170.3 hours and when we combine that with segmenting the dataset we get another $\sim$ 145.4 hours. These numbers are based on interpolating the results of our algorithm based on image size and quantity. In total, the algorithm would take about 14 days to process the entire archive using the segmentation algorithm. Utilizing a hybrid pipeline approach with a classifier that operates on lower resolution data was proved to be $\sim$ 90$\%$ as accurate as our segmentation algorithm \textcolor{black}{(see Table \ref{tab:comparison})}. Since the classifier is about 15 $\times$ faster than the segmentation network and we evaluate images at 1/16 resolution with it, we process the entire archive in $\sim$21 hours.

\textcolor{black}{
While our methodology is quick, it also needs to be trustworthy and accurate. It is too time consuming for us to manually evaluate every image that comes out of our pipeline however we have the capacity to inspect the positive detections since they're less numerous. Even though both algorithms flagged a handful of images we manually vetted a portion of them to ensure we had at least some reliable results to follow-up. Our vetting process consisted of manually inspecting the full-resolution images with the segmentation mask for brain terrain overlaid. We manually inspected 456 positive detections and rejected 255 as false positives. We can use the rejections to estimate a false positive rate of $\sim$0.4$\%$ (255/58209) and that value is remarkably similar to our estimated value in Table \ref{tab:comparison} which comes from a more idealized dataset. This outcome highlights the importance of encapsulating a diverse range of images in the training data in order to minimize false positives and ensure accurate results. One caveat about training on false positives (see section 2 about our active learning process) is that they can sometimes adversely effect the accuracy if the features are similar to brain terrain and there aren't many positive samples. We don't have a good prescription on how to systematically test this other than inspecting the image embeddings or using some form of hold-out validation during training.
}


The implications of identifying the processes responsible for the relatively young brain terrains on Mars are significant. \textcolor{black}{The confirmed brain terrain detections are centered primarily around 40N (see Figure \ref{fig:mars_latitude}). This preferential distribution in latitude can provide important clues about the climate conditions necessary for the formation of brain terrain on Mars. A majority of the detections in the southern hemisphere centered around 40S latitude and are classified as candidates (images flagged by both our classifier and segmentation network) because they require manual confirmation. Due to the numerous candidate detections we didn't have time to check them by hand and are releasing an interactive table on GitHub to help with labelling the $\sim$1100 candidates\footnote{https://github.com/pearsonkyle/Mars-Brain-Coral-Network}. A future study by Noe et al. (\textit{in prep.}) will explore correlations between the brain terrain locations and climate models when Mars had a different obliquity. This preferential distribution suggests the formation of brain terrain may be correlated with specific climate conditions, such as those related to past changes in Mars' obliquity. If brain terrain formed through freeze-thaw processes similar to patterned stone circles on Earth, it would imply that cryoturbation has redistributed rock near the Martian surface and that it is largely confined to a narrow latitudinal bands. Integrating multiple remote sensing datasets, including high-resolution imagery, crater counts, thermal data, and rock counts, can provide robust tests of the competing hypotheses for brain terrain formation.}

\section{Conclusion}

We developed a hybrid pipeline using convolutional neural networks to automate the detection of rare brain terrain in a large dataset of 58,209 HiRISE images of Mars. A Fourier domain classifier on compressed imagery allowed rapid screening before pixel-level segmentation mapping of landform boundaries, enabling processing of the 28TB archive in just 21 hours rather than what would have been 14 days. \textcolor{black}{We identify 201 new images containing brain terrain with an additional 1141 candidates that will require vetting by hand to be confirmed.} An interactive table of the results is available on Github for the community to explore and build upon\footnote{https://github.com/pearsonkyle/Mars-Brain-Coral-Network}. \textcolor{black}{The confirmed brain terrain detections are centered primarily around 40N with another small group around 40S. The preferential distribution in latitude can correlate to certain climate conditions based on Mars' history but more work is required to quantify the correlation exactly. If brain terrain formed by freeze-thaw cycles similar to the processes that create patterned stone circles on Earth (\citealt{Kessler2001}; \citealt{Kaab2014}), it would imply that freeze-thaw processes redistribute rocks near the martian surface and that this process is largely limited to a narrow latitudinal band. In a future study we will map rock distributions in HiRISE images using shadow measurements to derive size and abundance similar to \citealt{Golombek2012}; and \citealt{Huertas2006}. Quantifying rock distributions will help test hypotheses related to possible cryoturbation and freeze-thaw processes in brain terrain formation. Integrating multiple remote sensing datasets, such as high-resolution imagery, crater counts, thermal data, and rock counting from shadow analysis, can provide robust tests of the competing brain terrain formation hypotheses. This, in turn, can reveal valuable insights into Mars' recent climate, geology, and potential habitability, which have important implications for future landing site selection and the study of these rare Martian landforms.} \citep{Gallagher2011}

\section{Acknowledgements}

The research described in this publication was carried out in part at the Jet Propulsion Laboratory, California Institute of Technology, under a contract with the National Aeronautics and Space Administration. This research has made use of the High Resolution Imaging Science Experiment on the Mars Reconnaissance Orbiter, under contract with the National Aeronautics and Space Administration. We acknowledge funding support from the National Aeronautics and Space Administration (NASA) Mars Data Analysis Program (MDAP) Grant Number NNH19ZDA001N.

\bibliographystyle{aasjournal}
\bibliography{ref}

\end{document}